\documentclass{article}

\usepackage{microtype}
\usepackage{graphicx}
\usepackage{subcaption}
\usepackage{booktabs} 

\usepackage{hyperref}

\usepackage[preprint]{icml2026}

\usepackage{amsmath}
\usepackage{amssymb}
\usepackage{mathtools}
\usepackage{amsthm}

\usepackage[capitalize,noabbrev]{cleveref}

\theoremstyle{plain}

\theoremstyle{definition}

\theoremstyle{remark}

\usepackage[textsize=tiny]{todonotes}

\icmltitlerunning{}

\usepackage{xspace}
\usepackage{pifont}
\newcommand{\ours}{\textbf{GRACE-MoE}\xspace}

\begin{document}

\twocolumn[
  \icmltitle{GRACE-MoE: Grouping and Replication with Locality-Aware Routing \\ for Efficient Distributed MoE Inference}



  \icmlsetsymbol{equal}{*}

  \begin{icmlauthorlist}
    \icmlauthor{Yu Han}{USTC}
    \icmlauthor{Lehan Pan}{USTC}
    \icmlauthor{Jie Peng}{USTC}
    \icmlauthor{Ziyang Tao}{USTC}
    \icmlauthor{Hanqi Zhu}{USTC}
    \icmlauthor{Wuyang Zhang}{USTC}
    \icmlauthor{Yanyong Zhang}{USTC}
  \end{icmlauthorlist}

  \icmlaffiliation{USTC}{University of Science and Technology of China}
  
  \icmlcorrespondingauthor{Wuyang Zhang}{wuyangz@ustc.edu.cn}
  \icmlcorrespondingauthor{Yanyong Zhang}{yanyongz@ustc.edu.cn}


  \vskip 0.3in
]



\printAffiliationsAndNotice{}  

\begin{abstract}
  Sparse Mixture of Experts (SMoE) enables scalable parameter growth in large language models (LLMs) by selectively activating a subset of experts, and its large parameter count necessitates distributed deployment for inference. However, distributed inference faces a critical dilemma: although communication overhead constitutes the primary bottleneck, reducing it often exacerbates computational load imbalance, leading to resource waste. In this paper, we present~\ours, which stands for \textbf{\underline{G}}rouping and \textbf{\underline{R}}eplic\textbf{\underline{a}}tion with Lo\textbf{\underline{c}}ality-Awar\textbf{\underline{e}} Routing for S\textbf{\underline{MoE}} inference. \ours is a lossless co-optimization framework that integrates expert grouping to reduce communication and dynamic replication to correct load skew, together with locality-aware routing to resolve replica selection. To underpin this coordinated optimization in multi-node settings, \ours adopts a hierarchical sparse communication design that reduces cross-node traffic while implicitly aligning execution across nodes, thereby mitigating synchronization overhead. Experiments on diverse models and multi-node, multi-GPU environments demonstrate that~\ours efficiently reduces end-to-end inference latency, achieving up to \textbf{4.66$\times$} speedup over existing systems, and the code will be released upon acceptance.
\end{abstract}

\begin{figure*}[t]
  \centering 
    \begin{subfigure}[t]{0.48\textwidth}
      \centering
      \includegraphics[width=\linewidth]{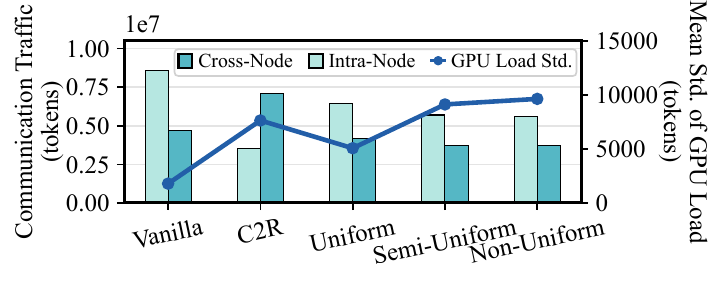}
      \caption{Uniformity constraint vs. communication traffic}
      \label{fig:grouping}
    \end{subfigure}
    \hfill
    \begin{subfigure}[t]{0.48\textwidth}
      \centering
      \includegraphics[width=\linewidth]{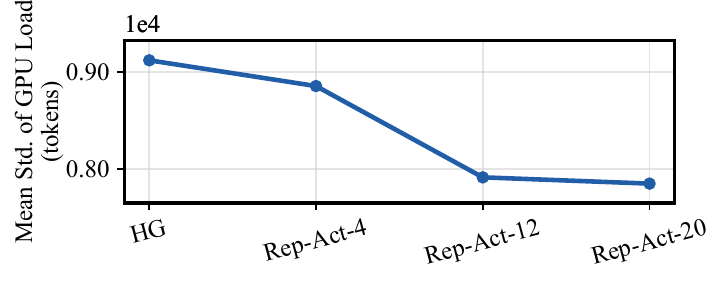}
       \caption{Number of replicated experts vs. computational load balance}
      \label{fig:replication}
    \end{subfigure}
    \caption{\textbf{Impact of grouping uniformity constraint and number of replicated experts.} Analysis of OLMoE with 2 nodes $\times$ 2 GPUs per node. Rep-Act-$x$ denotes replication of $x$ highly activated experts shared across HG groups; HG denotes hierarchical grouping.}
    \label{fig:observations}
\end{figure*}

\section{Introduction}
\label{sec:intro}
Large language models (LLMs) built on the Transformer architecture~\cite{vaswani2017attention} demonstrate substantial performance gains as parameter counts increase~\cite{brown2020language}, but scaling dense models by simply enlarging parameters incurs prohibitive computation and memory costs~\cite{kaplan2020scaling, clark2022unified}. 
The Sparse Mixture-of-Experts (SMoE) architecture mitigates this by partitioning parameters into experts and activating only a small subset per token, thereby enabling “large-parameter but small-computation” scaling~\cite{shazeer2017outrageously}.
Recent SMoE systems, such as GShard~\cite{lepikhin2021gshard} and Switch Transformer~\cite{fedus2022switch}, have reached trillion-parameter scales, underscoring this potential.

Unfortunately, the massive parameter scale of SMoE exceeds the memory and computation capacity of a single device (i.e., a GPU), necessitating distributed deployment with expert parallelism, combined with data parallelism~\cite{lepikhin2021gshard, zhai2023smartmoe}. 
In this setting, experts within each MoE layer are partitioned across GPUs and coordinated through All-to-All communication. This design introduces two critical bottlenecks for inference: communication overhead and computational load\footnote{\label{fn:load}Computational load refers to the number of tokens assigned to an expert, or the total over a group or GPU.} imbalance.

Each MoE layer involves two rounds of All-to-All communication, dispatching tokens to experts and aggregating results. 
Repeated across layers, this amplifies latency, making communication the primary bottleneck in SMoE inference~\cite{he2022fastermoe, gale2023megablocks}. 
In cross-node settings with limited bandwidth, All-to-All communication accounts for over 70\% of a single MoE layer's execution time and about 40\% of overall end-to-end inference latency across layers~\cite{li2023accelerating, hwang2023tutel}. 
Meanwhile, the gating network naturally skews token routing, creating “hot” and “cold” experts that cause load imbalance, overloading some GPUs while leaving others idle and wasting computing resources~\cite{lewis2021base, clark2022unified, he2022fastermoe}. 
Prior work typically addresses these two issues in isolation, but improving one often worsens the other. For example, methods that reduce communication often concentrate co-activated experts, which increases load skew. 
Such trade-offs can often be tolerated within a single node, where high-bandwidth links and tight synchronization partially mask their impact. 
However, in multi-node settings, limited cross-node bandwidth amplifies both effects: communication overhead becomes dominant, while load imbalance further exacerbates communication tail latency~\cite{go2025moetuner}, delaying global synchronization. 
As a result, jointly mitigating communication overhead and computational load imbalance in multi-node SMoE inference remains an open challenge. 

At the system level, this communication bottleneck manifests in how All-to-All communication is implemented in multi-node deployments. 
Most existing systems adopt a flat global All-to-All communication pattern that requires strict synchronization across all ranks within a communication group. 
In heterogeneous clusters where high-bandwidth intra-node links (e.g., NVLink) coexist with significantly slower cross-node links (e.g., Ethernet), global synchronization is often limited by the slowest links. 
This straggler effect substantially amplifies synchronization overhead and constitutes another critical scalability bottleneck for distributed multi-node, multi-GPU systems.

In this paper, we propose~\ours, a lossless co-optimization framework for multi-node SMoE inference. 
\ours consists of two tightly coupled phases: \ding{172} \textit{Grouping \& Replication} and \ding{173} \textit{Routing}, performed in the offline and online phases. 
During the offline phase, we group experts based on affinity (i.e., co-activation frequency) to reduce cross-device\footnote{Cross-device communication includes both intra-node and cross-node cases.} All-to-All communication, directly mitigating the communication bottleneck. We further replicate highly activated experts from the most heavily loaded groups, with the number of replicas dynamically determined by load skew, to alleviate computational load imbalance without excessive redundancy. 
In the online phase, we design a topology-aware routing strategy to determine which replica executes the computation. This strategy prioritizes local replicas and employs weighted round-robin with load prediction across remote replicas when necessary, maintaining load balance while limiting cross-node traffic. 
To make such joint optimization effective in multi-node environments, we introduce a hierarchical sparse communication design that reduces cross-node synchronization overhead through implicit alignment of execution across nodes, mitigating long-tail latency.
Together, these designs reconcile the conflicting objectives of communication efficiency and load balancing under the constraints of multi-node SMoE inference. 
Experiments on various MoE models and multi-node, multi-GPU setups show that~\ours reduces communication latency and alleviates load imbalance without accuracy degradation, reducing end-to-end inference latency by up to 78.55\% compared to existing systems.
The main contributions of this work are summarized as follows:
\begin{itemize}
    \item \textbf{Understanding trade-offs in SMoE inference.} 
    We identify the inherent trade-off between communication efficiency and load balancing.
    \item \textbf{A lossless co-optimization framework.} 
    We propose a framework that jointly optimizes communication efficiency and load balance without accuracy loss.
    \item \textbf{Hierarchical sparse communication architecture.} 
    We introduce a physically global yet logically sparse communication scheme for multi-node, multi-GPU SMoE inference.
\end{itemize}

\begin{figure*}[t]
  \centering
  \includegraphics[width=\textwidth]{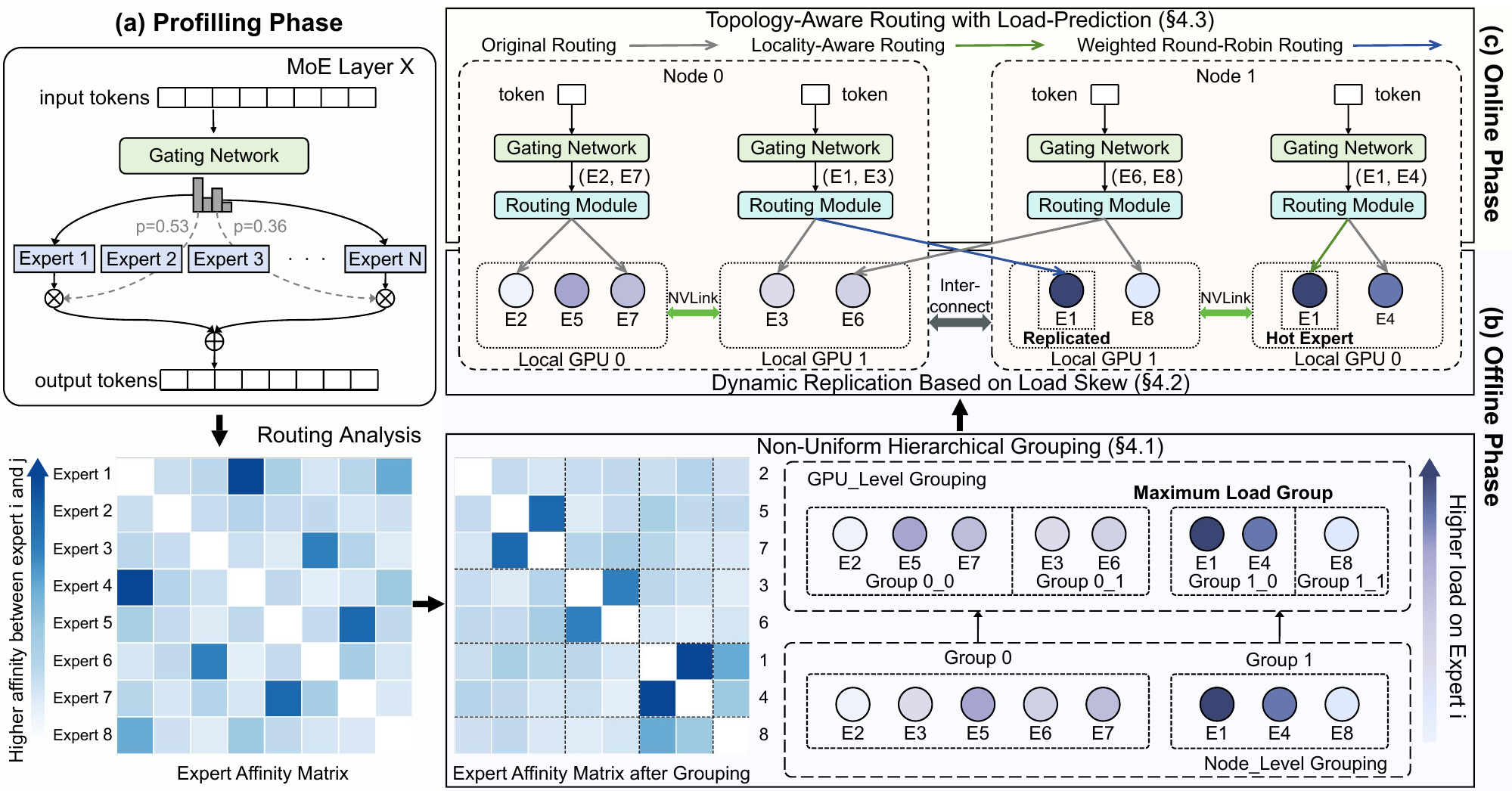}
  \caption{Overview of~\ours. (a) Profiling expert selections to build affinity matrices. (b) Grouping high-affinity experts on the same device and dynamically replicating hot experts to balance computational load. (c) Adaptive routing reduces communication by prioritizing local replicas and balances requests via weighted round-robin with load prediction across remote replicas.}
  \label{fig:overview}
\end{figure*}

\section{Related Work}
\textbf{Efficient SMoE Inference Systems.} 
Various systems have been proposed to accelerate SMoE inference, such as DeepSpeed-MoE~\cite{rasley2020deepspeed}, Tutel~\cite{hwang2023tutel}, and MegaBlocks~\cite{gale2023megablocks}, which optimize SMoE through kernel optimization, adaptive parallelism, and computation reformulation as block-sparse matrix multiplications. 
Further throughput improvements involve scheduling, memory management, and pipelining, as exemplified by Lina~\cite{li2023accelerating}, vLLM~\cite{kwon2023efficient}, Klotski~\cite{fang2025klotski}, and others~\cite{shen2022se, liu2023janus, huang2023towards, eliseev2023fast, zheng2024sglang, Kong2024SwapMoESO, li2024merge, wei2024aptmoe, yu2024moesys, hwang2024pre, Zhang2025CometFC, suo2025coserve}.
Despite these advances, communication overhead and load imbalance remain critical bottlenecks in expert-parallel inference.

\textbf{Expert Grouping, Replication and Routing.}
Some methods reduce communication through uniform expert grouping, such as C2R~\cite{zhang-etal-2025-advancing}, Occult~\cite{luo2025occult}, and others~\cite{yao2024exploiting, li2025speculative}. 
Among them, C2R and Occult achieve substantial communication savings for top-$k$ models but are lossy due to routing pruning and exacerbate load imbalance. 
Other works mitigate imbalance via expert replication or flexible placement~\cite{he2022fastermoe, nie2023flexmoe, wang2023prophet, wu2024lazarus, skiadopoulos2025accelerating, zeng2025efficientmoe, go2025moetuner, eplb_github}, yet most target training rather than inference and do not explicitly optimize communication.
Existing works mainly optimize either communication or load balance, while joint optimization remains largely unexplored, especially in multi-node settings.

\begin{figure*}[t]
  \centering
    \begin{subfigure}[t]{0.48\textwidth}
      \centering
      \includegraphics[width=\linewidth]{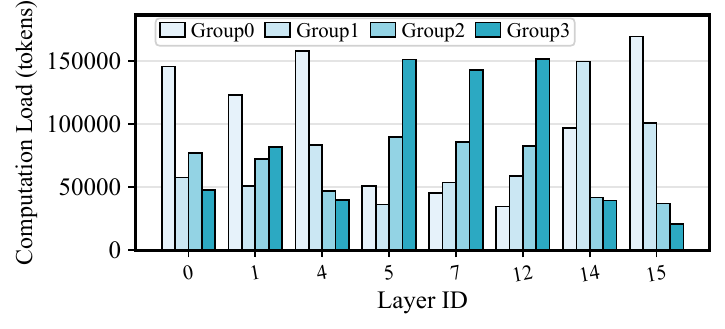}
      \caption{Group-level load analysis across layers}
      \label{fig:load-per-layer}
    \end{subfigure}
    \hfill
    \begin{subfigure}[t]{0.48\textwidth}
      \centering
      \includegraphics[width=\linewidth]{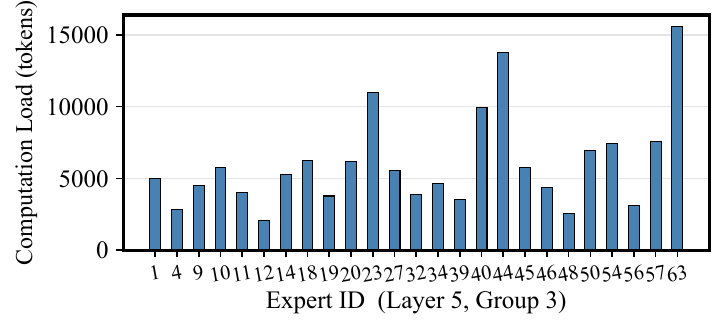}
      \caption{Per-expert load within the heaviest group}
      \label{fig:load-per-expert}
    \end{subfigure}
    \caption{\textbf{Computational load distribution after hierarchical grouping.} (a) In OLMoE, affinity clustering concentrates load on a few groups. (b) In Layer~5, the heaviest group’s per-expert load shows overload from a few frequently activated experts.}
    \label{fig:computational-load}
\end{figure*}

\section{Observations}
\label{sec:observation}
Motivated by the communication overhead and load imbalance discussed in Section~\ref{sec:intro}, we analyze communication traffic in intra-node and cross-node settings. 
Under top-$k$ routing, each token activates $k$ experts per layer, and C2R~\cite{zhang-etal-2025-advancing} shows that expert activations exhibit strong co-activation patterns. 
We define expert affinity as the frequency with which two experts are co-activated by the same tokens. 
Grouping experts by affinity naturally yields uneven group sizes.
Uniform grouping enforces equal experts per group, whereas non-uniform grouping relaxes this constraint and allows group sizes to follow affinity structure. 
As shown in Figure~\ref{fig:grouping}, relaxing the uniformity constraint better exploits affinity and reduces cross-device traffic compared to Vanilla and C2R, while uniform grouping disrupts co-activation and limits optimization. 
However, affinity-based grouping concentrates frequently co-activated experts into the same groups, increasing the chance that certain devices receive disproportionately more tokens and exacerbating computational load imbalance, especially under non-uniform grouping. 
This trade-off is evident in Figure~\ref{fig:grouping}, where grouping strategies that reduce communication traffic lead to worse load imbalance, motivating our design. 

Beyond grouping, communication patterns introduce additional challenges in multi-node settings. 
Aggregating token copies destined for the same node can reduce expensive cross-node bandwidth consumption, but this is unsupported by flat global All-to-All and motivates hierarchical All-to-All. 
However, conventional hierarchical implementations decompose communication into multiple stages, incurring extra kernel launches and synchronization overhead. 
More critically, physically partitioning communication groups across nodes without global coordination leads to progress decoupling, where faster groups contend more aggressively for cross-node bandwidth and force slower groups to stall, resulting in long-tail latency. 
This imbalance propagates to subsequent intra-node communication stages, causing GPU spin-waiting and amplified synchronization overhead.

\section{Method}
\label{sec:method}
To address the trade-off observed in Section~\ref{sec:observation}, we propose~\ours, a hybrid optimization framework built on profiling of routing behaviors.
During profiling, per-layer expert selections are recorded to construct expert affinity matrices and load statistics. 
Guided by this analysis, the framework integrates offline non-uniform hierarchical expert grouping (Section~\ref{sec:grouping}) and dynamic replication based on load skew (Section~\ref{sec:replication}) with online locality-aware routing with load prediction (Section~\ref{sec:routing policy}). 
As illustrated in Figure~\ref{fig:overview}, this comprehensive design effectively reduces communication overhead and improves computational load balance in multi-node, multi-GPU SMoE inference, while maintaining model accuracy.

\begin{figure*}[t]
    \centering
    \includegraphics[width=0.89\textwidth]{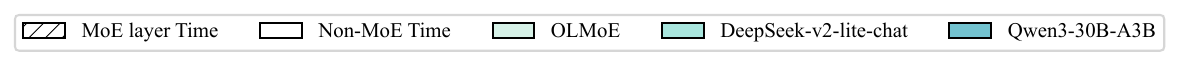}
    \begin{minipage}{0.49\textwidth}
        \centering\small 2 nodes $\times$ 2 GPUs/node
    \end{minipage}\hfill
    \begin{minipage}{0.49\textwidth}
        \centering\small 2 nodes $\times$ 4 GPUs/node
    \end{minipage}

    \begin{subfigure}{0.48\textwidth}
      \centering
      \includegraphics[width=\linewidth]{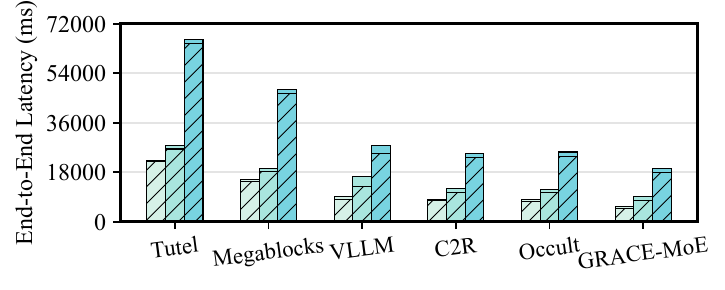}
      \caption{batch size = 256, prefill length = 128, decode length = 16}
    \end{subfigure}
    \hfill
    \begin{subfigure}{0.48\textwidth}
      \centering
      \includegraphics[width=\linewidth]{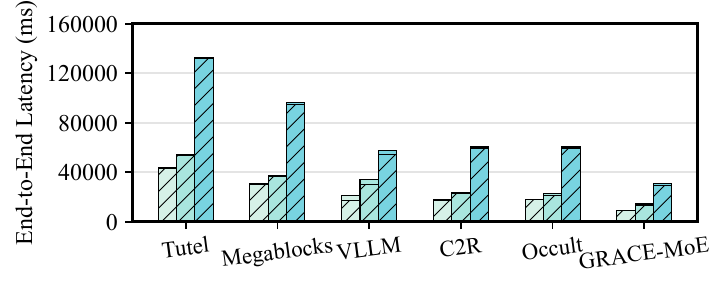}
      \caption{batch size = 256, prefill length = 128, decode length = 16}
    \end{subfigure}

    \begin{subfigure}{0.48\textwidth}
      \centering
      \includegraphics[width=\linewidth]{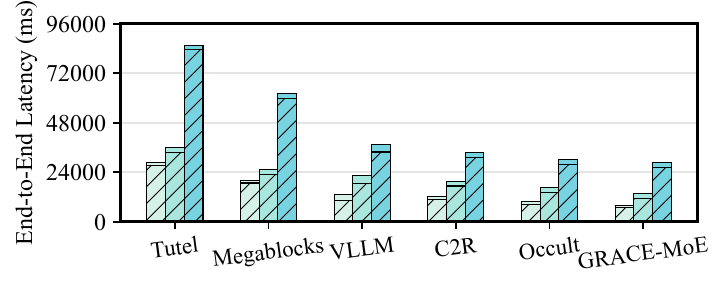}
      \caption{batch size = 512, prefill length = 64, decode length = 32}
    \end{subfigure}
    \hfill
    \begin{subfigure}{0.48\textwidth}
      \centering
      \includegraphics[width=\linewidth]{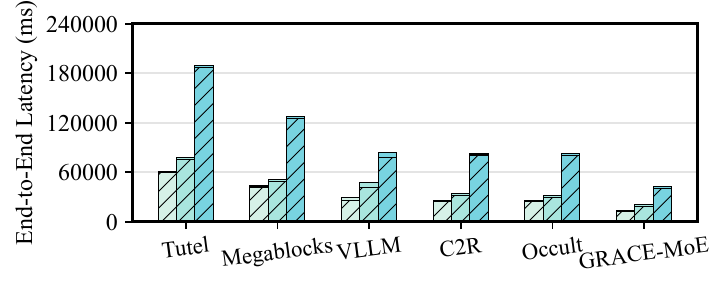}
      \caption{\small batch size = 512, prefill length = 64, decode length = 32}
    \end{subfigure}
    \caption{\textbf{End-to-end inference latency and MoE layer time.} Comparison of \ours and all baselines under different workloads and cluster settings.}
    \label{fig:e2e_1}
\end{figure*}

\subsection{Expert Grouping: Communication-Centric Optimization}
\label{sec:grouping}
The objective of expert grouping is to colocate high-affinity experts on the same GPU to reduce cross-device communication. 
We build on spectral clustering to design a hierarchical grouping scheme for multi-node, multi-GPU topologies. 

\textbf{Non-Uniform Grouping of Experts Based on Intra-Layer Affinity.} 
Spectral clustering produces groups with dense intra-connections and sparse inter-connections, aligning with our communication-centric goal.
As observed in Section~\ref{sec:observation}, affinity-based grouping tends to form uneven group sizes but better captures co-activation patterns, thereby reducing communication. 
We therefore apply spectral clustering to the expert affinity matrix to generate fully non-uniform groups, with sizes determined solely by affinity.
Although fully non-uniform grouping reduces communication, it leads to computational load imbalance that is even more severe than in the uniform scheme. To mitigate this, we propose controlled non-uniform grouping, regulated by a non-uniformity ratio $r$ that bounds group-size deviations. 
Given an ideal group size $E = \frac{n}{D}$, where $n$ is the number of experts per layer and $D$ is the number of groups, actual sizes are restricted to $[E-\delta, E+\delta]$, where $\delta = E \cdot r$. The choice of $r$ is critical: too small a value splits high-affinity experts and increases communication, while too large a value creates substantial group size disparity and worsens load imbalance. We model the selection of $r$ as an optimization problem that balances affinity utilization against grouping non-uniformity. We define intra-group affinity utilization $U(r)$ and size deviation $S(r)$ as
\begin{equation}
    U(r) = \frac{\sum_{C \in \mathcal{C}(r)} \sum_{\substack{i<j \\ i,j \in C}} A_{i,j}}{\sum_{i<j} A_{i,j}}.
\end{equation}
\begin{equation}
    S(r) = \sqrt{\frac{1}{D}\sum_{d=1}^{D}\left(\lvert C_d \rvert - {E}\right)^{2}}.
\end{equation}
where $r$ is the candidate ratio, $\mathcal{C}(r) = \{C_1,\dots,C_D\}$ denotes the grouping with $D$ groups, and $A \in \mathbb{R}^{n\times n}$ is the affinity matrix, with $A_{i,j}$ denoting the affinity between experts $i$ and $j$. By plotting $(S(r),U(r))$, we select the knee point as the optimal $r$, preserving affinity while avoiding excessive size gaps. 
The validity of this choice is empirically confirmed in Appendix~\ref{sec:NUR}. 
After determining $r$, we refine fully non-uniform grouping by reassigning experts with the lowest intra-group affinity to candidate groups with higher affinity, yielding a scheme with controlled non-uniformity. 
Details of the algorithm are provided in Appendix~\ref{sec:Alg.1}.

\textbf{Hierarchical Grouping for Distributed Expert Placement.} In multi-node, multi-GPU scenarios, we adopt a hierarchical grouping (HG) strategy. At the cross-node level, experts within each layer are divided into $N$ large groups mapped to nodes. Since cross-node communication is much more expensive, we apply fully non-uniform grouping to maximize intra-node affinity and minimize cross-node traffic. Within each node, these groups are further partitioned into $G$ smaller groups mapped to individual GPUs, where controlled non-uniform grouping is applied to balance group size while preserving affinity. This two-level strategy achieves communication optimization across the topology: affinity is maximized within GPUs, weaker across GPUs in the same node, and rare across nodes. As a result, communication overhead is significantly reduced.

\subsection{Expert Replication: Computational Load Balance-Centric Optimization}
\label{sec:replication}
The affinity-based expert grouping scheme reduces communication but also aggravates the inherent computational load imbalance of SMoE models. 
High-affinity experts are frequently co-activated, and when grouped together, they tend to overload their hosting GPU. 
To mitigate this imbalance while preserving the communication benefits of grouping, we propose dynamic expert replication.

\textbf{Selection of Experts for Replication.} 
The root cause of load imbalance in SMoE inference lies in a small number of experts being frequently activated. 
We therefore replicate highly activated experts. 
As shown in Figure~\ref{fig:replication}, starting from hierarchical grouping, we replicate different numbers of such experts by placing one replica on each GPU.
The results show that replicating only a few experts yields limited improvement, while a moderate replication level significantly reduces load imbalance.
However, further increasing the number of replicated experts brings only marginal additional benefits.
We attribute this to redundant replication, which degrades the system toward data parallelism, disrupts affinity-based grouping, and incurs unnecessary memory overhead.
Hence, the replication scope must be carefully constrained.
As illustrated in Figure~\ref{fig:computational-load}, after grouping, only a few groups in each layer handle the majority of tokens, and the overload mainly stems from a small number of frequently activated experts. We therefore replicate only these experts within the heaviest group rather than the entire group, preserving intra-group affinity and communication benefits while avoiding redundancy and wasted resources.

\textbf{Dynamic Replica Allocation Based on Load Skew.} 
Since expert activation distributions and grouping results vary across layers, the computational load skew of the heaviest group also differs. 
Therefore, we propose a dynamic replication (DR) strategy driven by load skew. 
After generating the expert groups in each layer, profiling data are used to calculate the load $W_i$ of each group, yielding the maximum $W_{\max}$ and mean load $\overline{W}$. The computational load skew factor ($\rho$) is defined as $W_{\max}/\overline{W}$, and the number of replicas is determined by~\cref{formula:n_replica}. 
\begin{equation}
    n_{\mathrm{replica}} = \min\left( \max\left(1, \left\lfloor \rho \right\rfloor \right),\; n_{\mathrm{gpu}} - 1 \right).
\label{formula:n_replica}
\end{equation}
Within the heaviest group, experts are ranked by individual load, and those whose cumulative load exceeds $W_{\max} \cdot \frac{n_{\mathrm{replica}}}{1 + n_{\mathrm{replica}}}$ are identified as hot experts. These experts are then replicated onto the $n_{\mathrm{replica}}$ most underutilized GPUs. The original primary replicas remain, while additional replicas serve only as secondary copies, keeping the grouping structure intact. This mechanism effectively redistributes the workload of hotspot GPUs while maintaining communication efficiency, significantly mitigating the imbalance amplified by grouping.

\subsection{Routing Policy: Co-Optimizes Communication and Computational Load}
\label{sec:routing policy}
After replication, multiple expert instances exist, and the system must decide which replica executes computation. The routing policy should balance two objectives: minimizing cross-device communication and balancing computational load. We explore two complementary strategies. 

\textbf{Weighted Round-Robin with Load Prediction.} 
After replication, each duplicated expert has $n_{\mathrm{replica}}+1$ instances distributed across different GPUs, and routing must decide which instance processes incoming tokens. To guide this decision, we leverage the pre-replication load statistics from Section~\ref{sec:replication} and predict the post-replication computational load of GPUs. Let $W_{\max}$ denote the pre-replication load of the heaviest group and $W_{\mathrm{r}}$ the total load of its replicated experts. Assuming this load is evenly split across all $n_{\mathrm{replica}}+1$ instances, the per-instance load is $W_{\mathrm{p}} = W_{\max}/(n_{\mathrm{replica}}+1)$. The updated loads are then:
\begin{equation}
    W_{\max}^{\prime} = W_{\max} - W_{\mathrm{r}} + W_{\mathrm{p}}, \quad
    W_{\mathrm{i}}^{\prime} = W_{\mathrm{i}} + W_{\mathrm{p}}.
\end{equation}
where $W_{\mathrm{i}}$ is the pre-replication load of a target replica-hosting GPU. 
Based on these predictions, routing weights are assigned inversely proportional to the predicted loads, and tokens are dispatched via the weighted round-robin (WRR) policy. 
This approach alleviates overload on hotspot GPUs by directing more tokens to less loaded GPUs. However, its inherent randomness can trigger unnecessary cross-device communication by routing tokens to remote GPUs even when local replicas exist. This limits effectiveness under high concurrency, especially in multi-node scenarios. 

\textbf{Topology-Aware Routing with Locality Preference.} 
In distributed clusters, communication overhead exhibits a clear hierarchy: intra-GPU communication incurs negligible overhead, followed by intra-node communication across GPUs, while cross-node communication is the most expensive. 
This hierarchy motivates a topology-aware routing (TAR) policy that prioritizes replicas based on physical locality. 
The scheme follows a hierarchical locality-first policy: (i) If a replica exists on the same GPU as the token, it is selected. (ii) Otherwise, a replica on another GPU within the same node is chosen. (iii) Only if no intra-node replica is available, is the token routed to a cross-node replica. Within each tier, if multiple replicas are available, weighted round-robin with load prediction is applied to balance computational load. While sacrificing some load balance, it significantly reduces communication overhead, which is the dominant bottleneck in large-scale inference, thereby achieving a practical trade-off between communication and computation. 
The details of our routing policies are provided in Appendix~\ref{sec:Alg.2}.

\section{System Design}
To address synchronization bottlenecks in multi-node, multi-GPU environments, we introduce hierarchical sparse communication (HSC) that replaces flat global All-to-All with a physically global but logically sparse communication scheme, enabling efficient token dispatch through a two-stage design. 
The first stage performs cross-node routing, where each GPU communicates with peer GPUs in remote nodes to forward tokens to target nodes, reducing cross-node traffic. 
The second stage redistributes tokens within each node to the GPUs hosting the target experts. 
Tokens routed to multiple experts on the same destination are transmitted only once.
Cross-node communication uses a single global communication group with zero-padding to realize logically sparse point-to-point transfers. 
This preserves the bandwidth benefits of sparse communication and leverages the implicit barrier of global collectives for soft synchronization across nodes.
Intra-node communication remains isolated to exploit high-bandwidth links and the scheduling flexibility of the hierarchical design. 
Cross-node communication is overlapped with intra-node routing decision computation via fine-grained pipelining, further reducing end-to-end latency.
Hierarchical sparse communication restructures costly cross-node transfers, mitigating long-tail latency and jitter while achieving stable end-to-end speedup.

\begin{table*}[t]
  \caption{\textbf{Component analysis.} Relative impact of incremental component optimizations under a 2 nodes $\times$ 2 GPUs/node setup. All values are reported as relative changes with respect to Occult, averaged over the three models.} 
  \label{tab:component_analysis}
  \centering
  \begin{small}
    \begin{sc}
      \setlength{\tabcolsep}{6pt}
      \renewcommand{\arraystretch}{1.08}
      \begin{tabular}{l|c|c|c|c|c|c}
        \toprule
        \textbf{Metric}
        & \textbf{Occult}
        & \textbf{Occult + HSC}
        & \textbf{HG + HSC}
        & \textbf{+ FR + WRR}
        & \textbf{+ DR + WRR}
        & \textbf{+ DR + TAR} \\
        \midrule
        All-to-All Time       & 0.00 & $-35.19\%$ & $-48.33\%$ & $-44.52\%$ & $-44.91\%$ & $-50.57\%$ \\
        Cross-node Traffic    & 0.00 & $-35.64\%$ & $-50.67\%$ & $-41.62\%$ & $-41.84\%$ & $-52.11\%$ \\
        Intra-node Traffic    & 0.00 & $+100.13\%$& $+47.10\%$ & $+76.33\%$ & $+78.41\%$ & $+57.22\%$ \\
        GPU Idle Time         & 0.00 & $-49.88\%$ & $-4.78\%$  & $-6.18\%$  & $-26.86\%$ & $-25.66\%$ \\
        Avg. GPU Load Std.    & 0.00 & $+0.02\%$  & $+90.03\%$ & $+51.32\%$ & $+31.92\%$ & $+39.35\%$ \\
        \bottomrule
      \end{tabular}
    \end{sc}
  \end{small}
\end{table*}

\section{Experiments}
\subsection{Experimental Setup}
\label{sec:setting}
\textbf{Models and Datasets.} 
We evaluate~\ours on three representative MoE models, OLMoE~\cite{muennighoff2024olmoe}, DeepSeek-v2-lite-chat~\cite{liu2024deepseek}, and Qwen3-30B-A3B~\cite{yang2025qwen3} (Table~\ref{table:config}). Datasets include WikiText-2-v1~\cite{merity2016pointer}, MATH~\cite{hendrycksmath2021}, and the GitHub subset of The Pile~\cite{gao2021pile}, covering text, code and math tasks.

\textbf{Baselines and Metrics.} 
Baselines include Tutel~\cite{hwang2023tutel}, Megablocks~\cite{gale2023megablocks}, vLLM~\cite{kwon2023efficient}, and expert placement methods C2R~\cite{zhang-etal-2025-advancing} and Occult~\cite{luo2025occult}. 
For Occult, we use its No-Prune variant as a lossless baseline. 
We measure communication efficiency, computational load balance, and inference performance using all-to-all communication time and traffic, GPU idle time, mean per-layer GPU load standard deviation, MoE layer time, and end-to-end latency.

\textbf{Hardware and Software.} 
Experiments are conducted on a multi-node system with two logical nodes, each equipped with 4$\times$ NVIDIA A100-SXM4 GPUs (80GB). GPUs within a node are interconnected via NVLink (12 links per GPU, 50\,GB/s per direction). 
Cross-node bandwidth is 25\,Gbps over Ethernet to emulate practical multi-node deployment. 
We implement~\ours on Megablocks~\cite{gale2023megablocks} using PyTorch 2.5~\cite{paszke2019pytorch} and Triton 3.1~\cite{tillet2019triton}, supporting data and expert parallelism. Inference uses BFloat16 precision. 

\subsection{End-to-End Performance}
\label{sec:e2e}
We evaluate~\ours on WikiText-2-v1~\cite{merity2016pointer} using the three MoE models in Section~\ref{sec:setting} under multi-node settings with 2 nodes $\times$ 2 GPUs/node and 2 nodes $\times$ 4 GPUs/node. 
To stress the system under diverse workloads, we vary the batch size and the ratio between prefill and decode lengths: (i) batch size = 256, prefill length = 128, decode length = 16; and (ii) batch size = 512, prefill length = 64, decode length = 32.
As shown in Figure~\ref{fig:e2e_1},~\ours consistently outperforms all baselines across models, workloads, and cluster scales, with the performance advantage becoming more pronounced at larger scales. 
While baseline systems exhibit steep latency growth due to rising cross-node communication overhead,~\ours effectively suppresses this trend, reducing MoE layer time by up to 80.11\%, 75.45\%, and 78.59\% and end-to-end latency by up to 78.55\%, 73.17\%, and 77.64\%, achieving maximum speedups of 4.66$\times$, 3.73$\times$, and 4.47$\times$ across the three models, respectively. 
These results demonstrate strong scalability.
Additional results in Appendix~\ref{sec:additional_e2e} show that~\ours maintains stable performance under lighter workloads on large clusters, highlighting robustness across deployment scenarios. 
Overall,~\ours improves end-to-end inference performance through the joint optimization of communication overhead and load balance in multi-node distributed settings, without loss of accuracy.

\subsection{Component Analysis}
\label{sec:component}
In multi-node SMoE inference, reducing communication overhead often aggravates load imbalance, while mitigating imbalance may increase communication. 
To study this trade-off, we use Occult~\cite{luo2025occult} as the uniform grouping baseline. 
We evaluate the three models from Section~\ref{sec:setting} on 2 nodes $\times$ 2 GPUs/node using WikiText-2-v1~\cite{merity2016pointer} under the workload (i) defined in Section~\ref{sec:e2e}.
We analyze communication efficiency, computational load balance, and their joint effect by decomposing MoE layer time into communication time, GPU idle time, and others. 

\textbf{Research Question 1: How to Reduce Communication Overhead?}
In distributed settings, All-to-All communication is the primary bottleneck. 
To mitigate it, we first introduce hierarchical sparse communication (HSC) while keeping Occult’s uniform expert placement. 
As shown in Table~\ref{tab:component_analysis}, HSC decreases All-to-All time by an average of 35.19\% and cross-node communication traffic by 35.64\% across the three models, while shifting traffic to intra-node links. 
Building on this system design, we further propose non-uniform hierarchical grouping (HG). 
Compared to Occult under the same system, HG shortens All-to-All time by 18.56\%, 17.96\%, and 24.69\% on the three models, and cuts cross-node communication traffic by 19.21\%, 16.72\%, and 34.77\%, respectively, while reducing intra-node traffic by 22.35\%, 15.00\%, and 42.15\%. 
Together, HSC and HG reshape communication patterns, substantially curbing cross-device transfers and improving communication efficiency in multi-node MoE inference.
\begin{figure}[h]
  \centering
    \centerline{\includegraphics[width=\columnwidth]{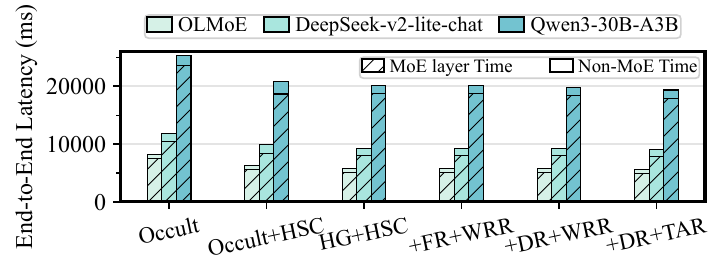}}
    \caption{\textbf{Ablation study of components.} Each variant corresponds to a component configuration reported in Table~\ref{tab:component_analysis}.} 
    \label{fig:ablation}
\end{figure}

\begin{figure*}[t]
    \centering
    \includegraphics[width=0.79\textwidth]{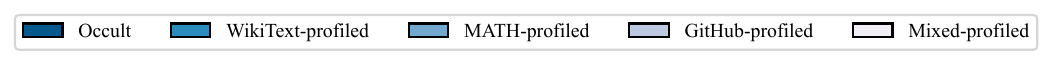}
    \begin{subfigure}{0.33\textwidth}
      \centering
      \includegraphics[width=\textwidth]{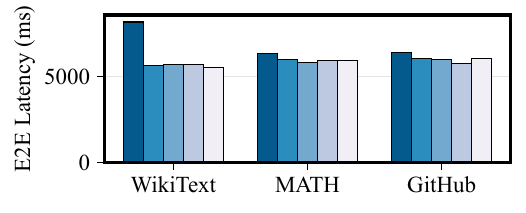}
      \caption{OLMoE}
      \label{fig:rq3_olmoe}
    \end{subfigure}
    \begin{subfigure}{0.33\textwidth}
      \centering
      \includegraphics[width=\textwidth]{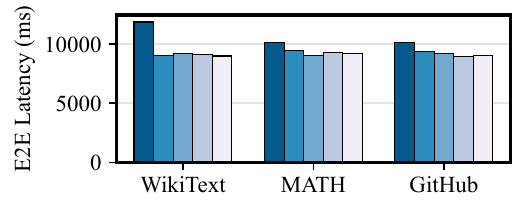}
      \caption{DeepSeek-v2-lite-chat}
      \label{fig:rq3_deepseek}
    \end{subfigure}
    \begin{subfigure}{0.33\textwidth}
      \centering
      \includegraphics[width=\textwidth]{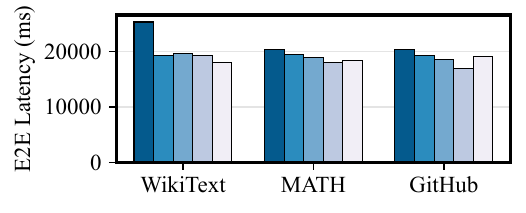}
      \caption{Qwen3-30B-A3B}
      \label{fig:rq3_qwen}
    \end{subfigure}
    \caption{\textbf{Generalization across datasets.} Placements derived from different datasets are cross-evaluated across three MoE models.}
    \label{fig:datasets}
\end{figure*}

\textbf{Research Question 2: How to Mitigate Computational Load Imbalance?}
As shown in Table~\ref{tab:component_analysis}, HG achieves the lowest communication overhead but exacerbates computational load imbalance. 
Compared to Occult under the same system design, GPU idle time increases by an average of 96.36\% across the three models, and the mean per-layer standard deviation of GPU load rises by 90.00\%, leaving some devices persistently underutilized. 
To address this issue, we introduce dynamic replication (DR) based on load skew, together with weighted round-robin routing with load prediction (WRR). 
In contrast to HG without replication, DR lowers GPU idle time by an average of 19.71\% and reduces GPU load deviation by 30.46\% across the three models, significantly improving utilization. 
For comparison, we evaluate a fixed-replica scheme (FR) that assigns one replica of overloaded experts in the heaviest group of each layer to the least-loaded GPU. 
This approach reduces GPU idle time by only 1.59\% on average, yielding limited improvements. 
Overall, the combination of DR and WRR achieves the lowest GPU idle time and the best load balance through adaptive replica allocation and weighted round-robin routing. 
Although replication increases the number of expert instances, replicas are created only for a small subset of heavily skewed experts per layer, keeping the parameter footprint within device memory limits.

\textbf{Research Question 3: How to Achieve Joint Optimization of Communication Overhead and Computational Load Balance?}
Table~\ref{tab:component_analysis} shows that DR with WRR mitigates load imbalance but introduces additional cross-device communication, increasing cross-node and intra-node traffic by 19.34\% and 23.20\% on average across the three models compared to HG without replication. 
To address this issue, we propose topology-aware routing with locality preference (TAR). 
Compared to WRR, TAR reduces All-to-All time by 9.47\%, 8.10\%, and 13.69\% on the three models, lowers cross-node communication traffic by 12.12\%, 12.42\%, and 29.40\%, and decreases intra-node traffic by an average of 12.58\%, while GPU idle time and load deviation increase only marginally by 2.58\% and 5.76\% on average. 
By prioritizing local replicas and falling back to WRR only when necessary, TAR achieves a more favorable trade-off between communication efficiency and computational load balance.

Finally, we evaluate how the components jointly translate to end-to-end performance. 
Figure~\ref{fig:ablation} shows end-to-end latency and MoE layer time as components are incrementally integrated and refined. 
Across all three models, progressively integrating hierarchical sparse communication, non-uniform hierarchical grouping, dynamic replication, and locality-aware routing consistently reduces both metrics. 
Compared to Occult, the full design achieves end-to-end speedups of 1.45$\times$, 1.31$\times$, and 1.31$\times$ on the three models, confirming that the component-level optimizations directly translate into practical inference gains. Absolute values of these metrics are shown in Appendix~\ref{sec:additional_component}.
\subsection{Generalizability Analysis}
Previous experiments demonstrate that~\ours generalizes well across different MoE models, workloads, and cluster scales. 
We further evaluate its generalizability under cross-dataset transfer. 
Expert grouping and replication schemes derived from profiling on individual datasets are directly applied to inference on other datasets under a 2 nodes $\times$ 2 GPUs/node setup with the workload (i) defined in Section~\ref{sec:e2e}. 
As shown in Figure~\ref{fig:datasets}, placements derived from one dataset consistently retain strong performance when transferred to others, despite distribution differences. 
Across three models and three target datasets, the worst-case end-to-end latency increase under cross-dataset placement is at most 4.52\% relative to in-domain placement, while still remaining at least 12.06\% lower than Occult~\cite{luo2025occult} on average across all evaluated model and dataset combinations. 
Placements derived from mixed-dataset profiling, combining samples from the three datasets mentioned in Section~\ref{sec:setting}, exhibit competitive and robust performance, often matching or even exceeding those obtained from single-dataset profiling. 
These results indicate that expert affinity and activation patterns captured by~\ours are stable across datasets, enabling reuse of offline grouping and replication without frequent re-profiling, thereby providing sustained inference performance gains across data distributions. 
This property is particularly important for practical deployment, where continuous online profiling would otherwise introduce additional overhead and system complexity.

\section{Conclusion}
We present~\ours, a co-optimization framework for jointly reducing communication overhead and alleviating computational load imbalance in distributed SMoE inference through non-uniform hierarchical grouping based on affinity, dynamic replication driven by load skew, and topology-aware routing with load prediction, supported by hierarchical sparse communication. 
Without sacrificing accuracy,~\ours improves end-to-end inference efficiency and demonstrates strong scalability, providing a practical solution for large-scale SMoE deployment.

\bibliography{example_paper}

@inproceedings{vaswani2017attention,
  author       = {Ashish Vaswani and Noam Shazeer and
                  Niki Parmar and
                  Jakob Uszkoreit and
                  Llion Jones and
                  Aidan N. Gomez and
                  Lukasz Kaiser and
                  Illia Polosukhin},
  editor       = {Isabelle Guyon and
                  Ulrike von Luxburg and
                  Samy Bengio and
                  Hanna M. Wallach and
                  Rob Fergus and
                  S. V. N. Vishwanathan and
                  Roman Garnett},
  title        = {Attention is All you Need},
  booktitle    = {Advances in Neural Information Processing Systems 30: Annual Conference
                  on Neural Information Processing Systems 2017, December 4-9, 2017,
                  Long Beach, CA, {USA}},
  pages        = {5998--6008},
  year         = {2017},
}

@article{kaplan2020scaling,
  title={Scaling laws for neural language models},
  author={Kaplan, Jared and McCandlish, Sam and Henighan, Tom and Brown, Tom B and Chess, Benjamin and Child, Rewon and Gray, Scott and Radford, Alec and Wu, Jeffrey and Amodei, Dario},
  journal={arXiv preprint arXiv:2001.08361},
  year={2020}
}

@inproceedings{clark2022unified,
  title={Unified scaling laws for routed language models},
  author={Clark, Aidan and de Las Casas, Diego and Guy, Aurelia and Mensch, Arthur and Paganini, Michela and Hoffmann, Jordan and Damoc, Bogdan and Hechtman, Blake and Cai, Trevor and Borgeaud, Sebastian and others},
  booktitle={International conference on machine learning},
  pages={4057--4086},
  year={2022},
  organization={PMLR}
}

@article{brown2020language,
  title={Language models are few-shot learners},
  author={Brown, Tom and Mann, Benjamin and Ryder, Nick and Subbiah, Melanie and Kaplan, Jared D and Dhariwal, Prafulla and Neelakantan, Arvind and Shyam, Pranav and Sastry, Girish and Askell, Amanda and others},
  journal={Advances in neural information processing systems},
  volume={33},
  pages={1877--1901},
  year={2020}
}

@inproceedings{shazeer2017outrageously,
  author       = {Noam Shazeer and
                  Azalia Mirhoseini and
                  Krzysztof Maziarz and
                  Andy Davis and
                  Quoc V. Le and
                  Geoffrey E. Hinton and
                  Jeff Dean},
  title        = {Outrageously Large Neural Networks: The Sparsely-Gated Mixture-of-Experts
                  Layer},
  booktitle    = {5th International Conference on Learning Representations, {ICLR} 2017,
                  Toulon, France, April 24-26, 2017, Conference Track Proceedings},
  publisher    = {OpenReview.net},
  year         = {2017},
  url          = {https://openreview.net/forum?id=B1ckMDqlg},
  bibsource    = {dblp computer science bibliography, https://dblp.org}
}

@inproceedings{lepikhin2021gshard,
  author       = {Dmitry Lepikhin and
                  HyoukJoong Lee and
                  Yuanzhong Xu and
                  Dehao Chen and
                  Orhan Firat and
                  Yanping Huang and
                  Maxim Krikun and
                  Noam Shazeer and
                  Zhifeng Chen},
  title        = {GShard: Scaling Giant Models with Conditional Computation and Automatic
                  Sharding},
  booktitle    = {9th International Conference on Learning Representations, {ICLR} 2021,
                  Virtual Event, Austria, May 3-7, 2021},
  publisher    = {OpenReview.net},
  year         = {2021},
  url          = {https://openreview.net/forum?id=qrwe7XHTmYb},
  bibsource    = {dblp computer science bibliography, https://dblp.org}
}

@article{fedus2022switch,
  title={Switch transformers: Scaling to trillion parameter models with simple and efficient sparsity},
  author={Fedus, William and Zoph, Barret and Shazeer, Noam},
  journal={Journal of Machine Learning Research},
  volume={23},
  number={120},
  pages={1--39},
  year={2022}
}

@inproceedings{zhai2023smartmoe,
  title={$\{$SmartMoE$\}$: Efficiently training $\{$Sparsely-Activated$\}$ models through combining offline and online parallelization},
  author={Zhai, Mingshu and He, Jiaao and Ma, Zixuan and Zong, Zan and Zhang, Runqing and Zhai, Jidong},
  booktitle={2023 USENIX Annual Technical Conference (USENIX ATC 23)},
  pages={961--975},
  year={2023}
}

@inproceedings{lewis2021base,
  title={Base layers: Simplifying training of large, sparse models},
  author={Lewis, Mike and Bhosale, Shruti and Dettmers, Tim and Goyal, Naman and Zettlemoyer, Luke},
  booktitle={International Conference on Machine Learning},
  pages={6265--6274},
  year={2021},
  organization={PMLR}
}

@inproceedings{rasley2020deepspeed,
  title={Deepspeed: System optimizations enable training deep learning models with over 100 billion parameters},
  author={Rasley, Jeff and Rajbhandari, Samyam and Ruwase, Olatunji and He, Yuxiong},
  booktitle={Proceedings of the 26th ACM SIGKDD international conference on knowledge discovery \& data mining},
  pages={3505--3506},
  year={2020}
}

@article{hwang2023tutel,
  title={Tutel: Adaptive mixture-of-experts at scale},
  author={Hwang, Changho and Cui, Wei and Xiong, Yifan and Yang, Ziyue and Liu, Ze and Hu, Han and Wang, Zilong and Salas, Rafael and Jose, Jithin and Ram, Prabhat and others},
  journal={Proceedings of Machine Learning and Systems},
  volume={5},
  pages={269--287},
  year={2023}
}

@article{gale2023megablocks,
  title={Megablocks: Efficient sparse training with mixture-of-experts},
  author={Gale, Trevor and Narayanan, Deepak and Young, Cliff and Zaharia, Matei},
  journal={Proceedings of Machine Learning and Systems},
  volume={5},
  pages={288--304},
  year={2023}
}

@inproceedings{liu2023janus,
  title={Janus: A unified distributed training framework for sparse mixture-of-experts models},
  author={Liu, Juncai and Wang, Jessie Hui and Jiang, Yimin},
  booktitle={Proceedings of the ACM SIGCOMM 2023 Conference},
  pages={486--498},
  year={2023}
}

@inproceedings{li2023accelerating,
  title={Accelerating distributed $\{$MoE$\}$ training and inference with lina},
  author={Li, Jiamin and Jiang, Yimin and Zhu, Yibo and Wang, Cong and Xu, Hong},
  booktitle={2023 USENIX Annual Technical Conference (USENIX ATC 23)},
  pages={945--959},
  year={2023}
}

@inproceedings{fang2025klotski,
  title={Klotski: Efficient Mixture-of-Expert Inference via Expert-Aware Multi-Batch Pipeline},
  author={Fang, Zhiyuan and Huang, Yuegui and Hong, Zicong and Lyu, Yufeng and Chen, Wuhui and Yu, Yue and Yu, Fan and Zheng, Zibin},
  booktitle={Proceedings of the 30th ACM International Conference on Architectural Support for Programming Languages and Operating Systems, Volume 2},
  pages={574--588},
  year={2025}
}

@inproceedings{li2024merge,
  author       = {Pingzhi Li and
                  Zhenyu Zhang and
                  Prateek Yadav and
                  Yi{-}Lin Sung and
                  Yu Cheng and
                  Mohit Bansal and
                  Tianlong Chen},
  title        = {Merge, Then Compress: Demystify Efficient SMoE with Hints from Its Routing Policy},
  booktitle    = {The Twelfth International Conference on Learning Representations,
                  {ICLR} 2024, Vienna, Austria, May 7-11, 2024},
  publisher    = {OpenReview.net},
  year         = {2024},
  url          = {https://openreview.net/forum?id=eFWG9Cy3WK},
  bibsource    = {dblp computer science bibliography, https://dblp.org}
}

@inproceedings{wei2024aptmoe,
  title={APTMoE: Affinity-Aware Pipeline Tuning for MoE Models on Bandwidth-Constrained GPU Nodes},
  author={Wei, Yuanxin and Du, Jiangsu and Jiang, Jiazhi and Shi, Xiao and Zhang, Xianwei and Huang, Dan and Xiao, Nong and Lu, Yutong},
  booktitle={SC24: International Conference for High Performance Computing, Networking, Storage and Analysis},
  pages={1--14},
  year={2024},
  organization={IEEE}
}

@article{yu2024moesys,
  title={Moesys: A distributed and efficient mixture-of-experts training and inference system for internet services},
  author={Yu, Dianhai and Shen, Liang and Hao, Hongxiang and Gong, Weibao and Wu, Huachao and Bian, Jiang and Dai, Lirong and Xiong, Haoyi},
  journal={IEEE Transactions on Services Computing},
  volume={17},
  number={5},
  pages={2626--2639},
  year={2024},
  publisher={IEEE}
}

@inproceedings{hwang2024pre,
  title={Pre-gated moe: An algorithm-system co-design for fast and scalable mixture-of-expert inference},
  author={Hwang, Ranggi and Wei, Jianyu and Cao, Shijie and Hwang, Changho and Tang, Xiaohu and Cao, Ting and Yang, Mao},
  booktitle={2024 ACM/IEEE 51st Annual International Symposium on Computer Architecture (ISCA)},
  pages={1018--1031},
  year={2024},
  organization={IEEE}
}

@article{shen2022se,
  title={Se-moe: A scalable and efficient mixture-of-experts distributed training and inference system},
  author={Shen, Liang and Wu, Zhihua and Gong, WeiBao and Hao, Hongxiang and Bai, Yangfan and Wu, HuaChao and Wu, Xinxuan and Bian, Jiang and Xiong, Haoyi and Yu, Dianhai and others},
  journal={arXiv e-prints},
  pages={arXiv--2205},
  year={2022}
}

@inproceedings{suo2025coserve,
  title={CoServe: Efficient Collaboration-of-Experts (CoE) Model Inference with Limited Memory},
  author={Suo, Jiashun and Liao, Xiaojian and Xiao, Limin and Ruan, Li and Wang, Jinquan and Su, Xiao and Huo, Zhisheng},
  booktitle={Proceedings of the 30th ACM International Conference on Architectural Support for Programming Languages and Operating Systems, Volume 2},
  pages={178--191},
  year={2025}
}

@inproceedings{kwon2023efficient,
  title={Efficient memory management for large language model serving with pagedattention},
  author={Kwon, Woosuk and Li, Zhuohan and Zhuang, Siyuan and Sheng, Ying and Zheng, Lianmin and Yu, Cody Hao and Gonzalez, Joseph and Zhang, Hao and Stoica, Ion},
  booktitle={Proceedings of the 29th symposium on operating systems principles},
  pages={611--626},
  year={2023}
}

@article{zheng2024sglang,
  title={Sglang: Efficient execution of structured language model programs},
  author={Zheng, Lianmin and Yin, Liangsheng and Xie, Zhiqiang and Sun, Chuyue Livia and Huang, Jeff and Yu, Cody Hao and Cao, Shiyi and Kozyrakis, Christos and Stoica, Ion and Gonzalez, Joseph E and others},
  journal={Advances in neural information processing systems},
  volume={37},
  pages={62557--62583},
  year={2024}
}

@inproceedings{yao2024exploiting,
  title={Exploiting inter-layer expert affinity for accelerating mixture-of-experts model inference},
  author={Yao, Jinghan and Anthony, Quentin and Shafi, Aamir and Subramoni, Hari and Panda, Dhabaleswar K DK},
  booktitle={2024 IEEE International Parallel and Distributed Processing Symposium (IPDPS)},
  pages={915--925},
  year={2024},
  organization={IEEE}
}

@article{go2025moetuner,
  title={Moetuner: Optimized mixture of expert serving with balanced expert placement and token routing},
  author={Go, Seokjin and Mahajan, Divya},
  journal={arXiv preprint arXiv:2502.06643},
  year={2025}
}

@inproceedings{zhang-etal-2025-advancing,
    title = "Advancing {M}o{E} Efficiency: A Collaboration-Constrained Routing ($\texttt{C2R}$) Strategy for Better Expert Parallelism Design",
    author = "Zhang, Mohan  and
      Li, Pingzhi  and
      Peng, Jie  and
      Qiu, Mufan  and
      Chen, Tianlong",
    editor = "Chiruzzo, Luis  and
      Ritter, Alan  and
      Wang, Lu",
    booktitle = "Proceedings of the 2025 Conference of the Nations of the Americas Chapter of the Association for Computational Linguistics: Human Language Technologies (Volume 1: Long Papers)",
    month = apr,
    year = "2025",
    address = "Albuquerque, New Mexico",
    publisher = "Association for Computational Linguistics",
    url = "https://aclanthology.org/2025.naacl-long.347/",
    doi = "10.18653/v1/2025.naacl-long.347",
    pages = "6815--6825",
    ISBN = "979-8-89176-189-6"
}

@article{li2025speculative,
  title={Speculative MoE: Communication Efficient Parallel MoE Inference with Speculative Token and Expert Pre-scheduling},
  author={Li, Yan and Zheng, Pengfei and Chen, Shuang and Xu, Zewei and Lai, Yuanhao and Du, Yunfei and Wang, Zhengang},
  journal={arXiv preprint arXiv:2503.04398},
  year={2025}
}

@article{skiadopoulos2025accelerating,
  title={Accelerating Mixture-of-Experts Training with Adaptive Expert Replication},
  author={Skiadopoulos, Athinagoras and Zhao, Mark and Gandhi, Swapnil and Norrie, Thomas and Mukherjee, Shrijeet and Kozyrakis, Christos},
  journal={arXiv preprint arXiv:2504.19925},
  year={2025}
}

@article{zeng2025efficientmoe,
  author       = {Yan Zeng and Chengchuang Huang and Yipeng Mei and Lifu Zhang and Teng Su and Wei Ye and Wenqi Shi and Shengnan Wang},
  title        = {EfficientMoE: Optimizing Mixture-of-Experts Model Training With Adaptive Load Balance},
  journal      = {{IEEE} Trans. Parallel Distributed Syst.},
  volume       = {36},
  number       = {4},
  pages        = {677--688},
  year         = {2025},
  url          = {https://doi.org/10.1109/TPDS.2025.3539297},
  doi          = {10.1109/TPDS.2025.3539297},
  bibsource    = {dblp computer science bibliography, https://dblp.org}
}

@inproceedings{he2022fastermoe,
  title={Fastermoe: modeling and optimizing training of large-scale dynamic pre-trained models},
  author={He, Jiaao and Zhai, Jidong and Antunes, Tiago and Wang, Haojie and Luo, Fuwen and Shi, Shangfeng and Li, Qin},
  booktitle={Proceedings of the 27th ACM SIGPLAN Symposium on Principles and Practice of Parallel Programming},
  pages={120--134},
  year={2022}
}

@article{wu2024lazarus,
  title={Lazarus: Resilient and elastic training of mixture-of-experts models with adaptive expert placement},
  author={Wu, Yongji and Qu, Wenjie and Tao, Tianyang and Wang, Zhuang and Bai, Wei and Li, Zhuohao and Tian, Yuan and Zhang, Jiaheng and Lentz, Matthew and Zhuo, Danyang},
  journal={arXiv preprint arXiv:2407.04656},
  year={2024}
}

@article{nie2023flexmoe,
  title={Flexmoe: Scaling large-scale sparse pre-trained model training via dynamic device placement},
  author={Nie, Xiaonan and Miao, Xupeng and Wang, Zilong and Yang, Zichao and Xue, Jilong and Ma, Lingxiao and Cao, Gang and Cui, Bin},
  journal={Proceedings of the ACM on Management of Data},
  volume={1},
  number={1},
  pages={1--19},
  year={2023},
  publisher={ACM New York, NY, USA}
}

@inproceedings{wang2023prophet,
  title={Prophet: Fine-grained load balancing for parallel training of large-scale moe models},
  author={Wang, Wei and Lai, Zhiquan and Li, Shengwei and Liu, Weijie and Ge, Keshi and Liu, Yujie and Shen, Ao and Li, Dongsheng},
  booktitle={2023 IEEE International Conference on Cluster Computing (CLUSTER)},
  pages={82--94},
  year={2023},
  organization={IEEE}
}

@inproceedings{muennighoff2024olmoe,
  author       = {Niklas Muennighoff and
                  Luca Soldaini and
                  Dirk Groeneveld and
                  Kyle Lo and
                  Jacob Morrison and
                  Sewon Min and
                  Weijia Shi and
                  Evan Pete Walsh and
                  Oyvind Tafjord and
                  Nathan Lambert and
                  Yuling Gu and
                  Shane Arora and
                  Akshita Bhagia and
                  Dustin Schwenk and
                  David Wadden and
                  Alexander Wettig and
                  Binyuan Hui and
                  Tim Dettmers and
                  Douwe Kiela and
                  Ali Farhadi and
                  et al.},
  title        = {OLMoE: Open Mixture-of-Experts Language Models},
  booktitle    = {The Thirteenth International Conference on Learning Representations,
                  {ICLR} 2025, Singapore, April 24-28, 2025},
  publisher    = {OpenReview.net},
  year         = {2025},
  url          = {https://openreview.net/forum?id=xXTkbTBmqq},
  bibsource    = {dblp computer science bibliography, https://dblp.org}
}

@article{liu2024deepseek,
  title={Deepseek-v2: A strong, economical, and efficient mixture-of-experts language model},
  author={Liu, Aixin and Feng, Bei and Wang, Bin and Wang, Bingxuan and Liu, Bo and Zhao, Chenggang and Dengr, Chengqi and Ruan, Chong and Dai, Damai and Guo, Daya and others},
  journal={arXiv preprint arXiv:2405.04434},
  year={2024}
}

@article{yang2025qwen3,
  title={Qwen3 technical report},
  author={Yang, An and Li, Anfeng and Yang, Baosong and Zhang, Beichen and Hui, Binyuan and Zheng, Bo and Yu, Bowen and Gao, Chang and Huang, Chengen and Lv, Chenxu and others},
  journal={arXiv preprint arXiv:2505.09388},
  year={2025}
}

@inproceedings{merity2016pointer,
  author       = {Stephen Merity and
                  Caiming Xiong and
                  James Bradbury and
                  Richard Socher},
  title        = {Pointer Sentinel Mixture Models},
  booktitle    = {5th International Conference on Learning Representations, {ICLR} 2017,
                  Toulon, France, April 24-26, 2017, Conference Track Proceedings},
  publisher    = {OpenReview.net},
  year         = {2017},
  url          = {https://openreview.net/forum?id=Byj72udxe},
  bibsource    = {dblp computer science bibliography, https://dblp.org}
}

@inproceedings{hendrycksmath2021,
  author       = {Dan Hendrycks and
                  Collin Burns and
                  Saurav Kadavath and
                  Akul Arora and
                  Steven Basart and
                  Eric Tang and
                  Dawn Song and
                  Jacob Steinhardt},
  editor       = {Joaquin Vanschoren and
                  Sai{-}Kit Yeung},
  title        = {Measuring Mathematical Problem Solving With the {MATH} Dataset},
  booktitle    = {Proceedings of the Neural Information Processing Systems Track on
                  Datasets and Benchmarks 1, NeurIPS Datasets and Benchmarks 2021, December
                  2021, virtual},
  year         = {2021},
}

@article{gao2021pile,
  author       = {Leo Gao and
                  Stella Biderman and
                  Sid Black and
                  Laurence Golding and
                  Travis Hoppe and
                  Charles Foster and
                  Jason Phang and
                  Horace He and
                  Anish Thite and
                  Noa Nabeshima and
                  Shawn Presser and
                  Connor Leahy},
  title        = {The Pile: An 800GB Dataset of Diverse Text for Language Modeling},
  journal      = {CoRR},
  volume       = {abs/2101.00027},
  year         = {2021},
  url          = {https://arxiv.org/abs/2101.00027},
  eprinttype    = {arXiv},
  eprint       = {2101.00027},
  bibsource    = {dblp computer science bibliography, https://dblp.org}
}

@inproceedings{paszke2019pytorch,
  author       = {Adam Paszke and
                  Sam Gross and
                  Francisco Massa and
                  Adam Lerer and
                  James Bradbury and
                  Gregory Chanan and
                  Trevor Killeen and
                  Zeming Lin and
                  Natalia Gimelshein and
                  Luca Antiga and
                  Alban Desmaison and
                  Andreas K{\"{o}}pf and
                  Edward Z. Yang and
                  Zachary DeVito and
                  Martin Raison and
                  Alykhan Tejani and
                  Sasank Chilamkurthy and
                  Benoit Steiner and
                  Lu Fang and
                  Junjie Bai and
                  Soumith Chintala},
  editor       = {Hanna M. Wallach and
                  Hugo Larochelle and
                  Alina Beygelzimer and
                  Florence d'Alch{\'{e}}{-}Buc and
                  Emily B. Fox and
                  Roman Garnett},
  title        = {PyTorch: An Imperative Style, High-Performance Deep Learning Library},
  booktitle    = {Advances in Neural Information Processing Systems 32: Annual Conference
                  on Neural Information Processing Systems 2019, NeurIPS 2019, December
                  8-14, 2019, Vancouver, BC, Canada},
  pages        = {8024--8035},
  year         = {2019},
}

@inproceedings{tillet2019triton,
  title={Triton: an intermediate language and compiler for tiled neural network computations},
  author={Tillet, Philippe and Kung, Hsiang-Tsung and Cox, David},
  booktitle={Proceedings of the 3rd ACM SIGPLAN International Workshop on Machine Learning and Programming Languages},
  pages={10--19},
  year={2019}
}

@inproceedings{luo2025occult,
  author       = {Shuqing Luo and
                  Pingzhi Li and
                  Jie Peng and
                  Yang Zhao and
                  Yu Cao and
                  Yu Cheng and
                  Tianlong Chen},
  title        = {Occult: Optimizing Collaborative Communications across Experts for
                  Accelerated Parallel MoE Training and Inference},
  booktitle    = {Forty-second International Conference on Machine Learning, {ICML}
                  2025, Vancouver, BC, Canada, July 13-19, 2025},
  publisher    = {OpenReview.net},
  year         = {2025},
  url          = {https://openreview.net/forum?id=vh2Dt4sT67},
  bibsource    = {dblp computer science bibliography, https://dblp.org}
}

@misc{eplb_github,
  author       = {DeepSeek-AI},
  title        = {\{EPLB\}: Expert Parallelism Load Balancer},
  howpublished = {\url{https://github.com/deepseek-ai/EPLB}},
  year         = {2025},
  note         = {Accessed: 2026-01-26}
}

@inproceedings{Zhang2025CometFC,
  author       = {Shulai Zhang and
                  Ningxin Zheng and
                  Haibin Lin and
                  Ziheng Jiang and
                  Wenlei Bao and
                  Chengquan Jiang and
                  Qi Hou and
                  Weihao Cui and
                  Size Zheng and
                  Li{-}Wen Chang and
                  Quan Chen and
                  Xin Liu},
  editor       = {Matei Zaharia and
                  Gauri Joshi and
                  Yingyan (Celine) Lin},
  title        = {{COMET:} Fine-grained Computation-communication Overlapping for Mixture-of-Experts},
  booktitle    = {Proceedings of the Eighth Conference on Machine Learning and Systems,
                  MLSys 2025, Santa Clara, CA, USA, May 12-15, 2025},
  publisher    = {OpenReview.net/mlsys.org},
  year         = {2025},
  url          = {https://openreview.net/forum?id=fGgQS5VW09},
  bibsource    = {dblp computer science bibliography, https://dblp.org}
}

@inproceedings{Kong2024SwapMoESO,
  author       = {Rui Kong and
                  Yuanchun Li and
                  Qingtian Feng and
                  Weijun Wang and
                  Xiaozhou Ye and
                  Ye Ouyang and
                  Linghe Kong and
                  Yunxin Liu},
  editor       = {Lun{-}Wei Ku and
                  Andre Martins and
                  Vivek Srikumar},
  title        = {SwapMoE: Serving Off-the-shelf MoE-based Large Language Models with
                  Tunable Memory Budget},
  booktitle    = {Proceedings of the 62nd Annual Meeting of the Association for Computational
                  Linguistics (Volume 1: Long Papers), {ACL} 2024, Bangkok, Thailand,
                  August 11-16, 2024},
  pages        = {6710--6720},
  publisher    = {Association for Computational Linguistics},
  year         = {2024},
  url          = {https://doi.org/10.18653/v1/2024.acl-long.363},
  doi          = {10.18653/V1/2024.ACL-LONG.363},
  bibsource    = {dblp computer science bibliography, https://dblp.org}
}

@article{huang2023towards,
  title={Towards moe deployment: Mitigating inefficiencies in mixture-of-expert (moe) inference},
  author={Huang, Haiyang and Ardalani, Newsha and Sun, Anna and Ke, Liu and Lee, Hsien-Hsin S and Sridhar, Anjali and Bhosale, Shruti and Wu, Carole-Jean and Lee, Benjamin},
  journal={arXiv preprint arXiv:2303.06182},
  year={2023}
}

@article{eliseev2023fast,
  title={Fast inference of mixture-of-experts language models with offloading},
  author={Eliseev, Artyom and Mazur, Denis},
  journal={arXiv preprint arXiv:2312.17238},
  year={2023}
}
\bibliographystyle{icml2026}

\newpage
\appendix
\onecolumn
\section{Appendix}



\subsection{Validation of Non-Uniform Ratio Selection}
\label{sec:NUR}
We conduct experiments using OLMoE~\citep{muennighoff2024olmoe} and WikiText-2-v1~\citep{merity2016pointer} under a 2 nodes $\times$ 2 GPUs/node setting to validate the selection of the non-uniform ratio ($r$). 
We compare three representative placements: uniform grouping (as used by Occult~\cite{luo2025occult}), fully non-uniform grouping, and controlled non-uniform grouping with $r=0.15$.
As shown in Table~\ref{table:nur}, uniform grouping suffers from higher All-to-All communication overhead due to disrupted expert affinity. 
In contrast, fully non-uniform grouping reduces communication but introduces severe computational load skew.
Although it achieves only marginal additional reduction in All-to-All time compared to controlled non-uniformity, it substantially increases GPU idle time, resulting in higher end-to-end latency.
Controlled non-uniform grouping with $r=0.15$ achieves a better balance between extremes, reducing communication overhead while limiting load imbalance. 
Overall, these results demonstrate the effectiveness of our non-uniform ratio ($r$) selection scheme.
\begin{table}[h]
    \caption{Comparison of different expert grouping strategies}
    \label{table:nur}
    \centering
    \resizebox{\columnwidth}{!}{
        \begin{sc}
            \begin{tabular}{lccc}
            \toprule
            \textbf{Grouping} & \textbf{All-to-All Time (ms)} & \textbf{GPU Idle Time (ms)} & \textbf{End-to-End Latency (ms)} \\
            \midrule
             Uniform (Occult)                       & 3494.02 & 501.69 & 6328.03 \\
             Controlled  Non-uniform($r=0.15$)      & 2846.38 & 506.93 & 5698.10 \\
             Fully Non-uniform                      & 2825.81 & 617.16 & 5747.50 \\
            \bottomrule
            \end{tabular}
        \end{sc}
    }    
\end{table}

\subsection{Algorithm for Controlled Non-uniform Grouping}
\label{sec:Alg.1}
For completeness, we provide the detailed pseudocode of our grouping scheme. 
Algorithm~\ref{alg:intrascore} defines the intra-group affinity score used to evaluate candidate assignments, while Algorithm~\ref{alg:CNUgrouping} gives 
the full procedure for controlled non-uniform grouping with non-uniformity ratio $r$. 
\begin{algorithm}[h]
    \caption{Intra-group Affinity Score}
    \label{alg:intrascore}
        \begin{algorithmic}
            \STATE {\bfseries Input:} affinity matrix $A \in \mathbb{R}^{n\times n}$, expert set $S$
            \STATE {\bfseries Output:} intra-group affinity score
            \STATE $score \leftarrow \sum_{i \in S} \sum_{j \in S} A_{i,j}$
            \STATE {\bfseries Return:} $score$
        \end{algorithmic}
\end{algorithm}
\begin{algorithm}[h]
    \caption{Controlled Non-uniform Grouping}
    \label{alg:CNUgrouping}
    \begin{algorithmic}
        \STATE {\bfseries Input:} affinity matrix $A \in \mathbb{R}^{n\times n}$, number of groups $D$, ratio $r$, experts $N_e$
        \STATE {\bfseries Output:} grouping $\{L_d\}_{d=1}^D$
        \STATE $E \leftarrow \left\lfloor N_e / D \right\rfloor$
        \STATE $\delta \leftarrow \max(1,\mathrm{round}(E \cdot r))$
        \STATE $num_{\min} \leftarrow \max(1, E-\delta)$
        \STATE $num_{\max} \leftarrow E+\delta$
        \STATE Initialize $L \leftarrow \{L_1,\dots,L_D\}$ as empty groups
        \STATE $\{C_d\}_{d=1}^D \leftarrow \textsc{SpectralClustering}(A, D)$
        \STATE $\Omega \leftarrow \emptyset$
        
        \FOR{$d=1$ {\bfseries to} $D$}
          \IF{$|C_d| > num_{\max}$}
            \STATE Keep top-$num_{\max}$ experts in $L_d$ by affinity, push others to $\Omega$
          \ELSE
            \STATE $L_d \leftarrow C_d$
          \ENDIF
        \ENDFOR
        
        \FOR{{\bfseries each} $e \in \Omega$}
          \STATE Assign $e$ to group $d^\star$ that maximizes intra-group affinity
          \STATE $L_{d^\star} \leftarrow L_{d^\star} \cup \{e\}$
        \ENDFOR
        
        \STATE Compute $need[d] \leftarrow \max(0,\, num_{\min} - |L_d|)$ for $d=1..D$
        \STATE $S \leftarrow \sum_{d=1}^D need[d]$
        
        \IF{$S > 0$}
          \STATE Move weakest-affinity experts from oversized groups to needy groups
        \ENDIF
        
        \STATE {\bfseries Return:} $\{L_d\}_{d=1}^D$
    \end{algorithmic}
\end{algorithm}

\subsection{Algorithm for Topology-Aware Routing with Load Prediction}
\label{sec:Alg.2}
We present two routing policies for replica assignment. Algorithm~\ref{alg:wrr} specifies the weighted polling strategy, while Algorithm~\ref{alg:topology} incorporates it into a topology-aware routing policy. Together, these ensure that replicas are selected with minimal cross-device communication overhead while maintaining balanced computational load.
\begin{algorithm}[h]
\caption{Weighted Round-Robin with Load Prediction}
\label{alg:wrr}
    \begin{algorithmic}
        \STATE {\bfseries Input:} polling weights (map gpu\_id $\rightarrow$ weight)
        \STATE {\bfseries Output:} selected\_gpu\_id
        \STATE $gpus \leftarrow \textsc{Keys}(\textit{polling\_weights})$
        \STATE $weights \leftarrow \textsc{Values}(\textit{polling\_weights})$
        \STATE $selected\_gpu\_id \leftarrow \textsc{WeightedRandomChoice}(gpus, weights)$
        \STATE {\bfseries return} $selected\_gpu\_id$
    \end{algorithmic}
\end{algorithm}

\begin{algorithm}[h]
\caption{Topology-Aware Routing with Locality Preference}
\label{alg:topology}
    \begin{algorithmic}
        \STATE {\bfseries Input:} token\_gpu\_id, token\_node\_id; replica\_gpus; polling\_weights (gpu\_id $\rightarrow$ weight)
        \STATE {\bfseries Output:} selected gpu\_id
        
        \STATE $local\_gpu\_replicas \leftarrow \{\, g \in replica\_gpus \mid g = token\_gpu\_id \,\}$
        \STATE $local\_node\_replicas \leftarrow \{\, g \in replica\_gpus \mid \textsc{Node}(g) = token\_node\_id \,\}$
        
        \IF{$local\_gpu\_replicas \neq \emptyset$}
          \STATE {\bfseries return} $token\_gpu\_id$
        \ELSE
          \IF{$local\_node\_replicas \neq \emptyset$}
            \STATE $local\_weights \leftarrow$ polling\_weights restricted to local\_node\_replicas
            \STATE {\bfseries return} $\textsc{ChooseByPollingWeight}(local\_weights)$
          \ELSE
            \STATE {\bfseries return} $\textsc{ChooseByPollingWeight}(polling\_weights)$
          \ENDIF
        \ENDIF
    \end{algorithmic}
\end{algorithm}

\subsection{Experiment Configurations}
\label{sec:config}
Details of the model architectures used for evaluation are summarized in Table~\ref{table:config}.  
\begin{table}[h]
    \caption{Model architecture details used in experiments}
    \label{table:config}
    \centering
        \begin{small}
        \begin{sc}
            \begin{tabular}{lcccccc}
            \toprule
            \textbf{Model} & 
            \textbf{Top\_k} & 
            \textbf{Experts} &
            \textbf{MoE Layers} &
            \textbf{Params} \\
            \midrule
                OLMoE                   & 8 & 64 & 16 & 6.92B \\
                DeepSeek-v2-lite-chat   & 6 & 64 & 26 & 15.7B\\
                Qwen3-30B-A3B           & 8 & 128 & 48 & 30.5B\\
            \bottomrule
            \end{tabular}
        \end{sc}
        \end{small}    
\end{table}

\subsection{Additional End-to-End Performance Results}
\label{sec:additional_e2e}
\begin{figure*}[t]
    \centering
    \includegraphics[width=0.89\linewidth]{figs/fig4/4_legend.pdf}
    \begin{subfigure}{0.48\textwidth}
      \centering
      \includegraphics[width=\linewidth]{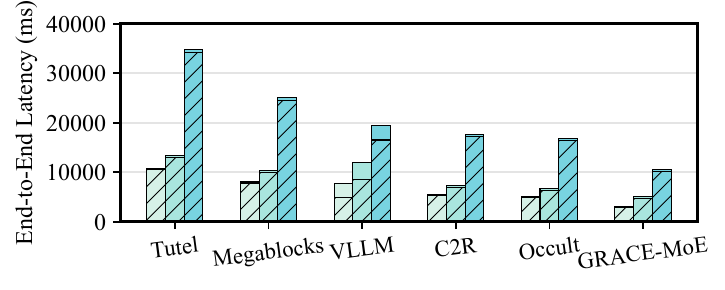}
      \caption{\small batch size = 64, prefill length = 128, decode length = 16}
    \end{subfigure}
    \hfill
    \begin{subfigure}{0.48\textwidth}
      \centering
      \includegraphics[width=\linewidth]{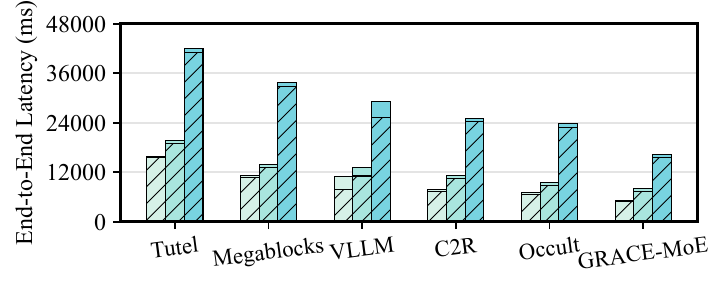}
      \caption{\small batch size = 128, prefill length = 64, decode length = 32}
    \end{subfigure}
    \caption{\textbf{End-to-end inference latency and MoE layer time under lighter workloads.} Supplementary evaluation of~\ours and all baselines across three models under the 2 nodes $\times$ 4 GPUs/node setting.}
    \label{fig:e2e_2}
\end{figure*}

To complement the main results in Section~\ref{sec:e2e}, we report additional end-to-end performance under two lighter workload configurations: (i) batch size = 64, prefill = 128, decode = 16; and (ii) batch size = 128, prefill = 64, decode = 32, evaluated on 2 nodes $\times$ 4 GPUs/node setting.
As shown in Figure~\ref{fig:e2e_2}, the performance trends remain consistent with those observed in Figure~\ref{fig:e2e_1}. Across both workloads,~\ours achieves lower end-to-end latency and shorter MoE layer time than all baselines. 
Notably, even under lighter workloads where communication pressure is reduced,~\ours consistently maintains performance advantages, demonstrating robust effectiveness across a wide range of workload intensities.

\subsection{Additional Component Analysis Results}
\label{sec:additional_component}
\begin{figure*}[t]
  \centering
    \includegraphics[width=0.55\textwidth]{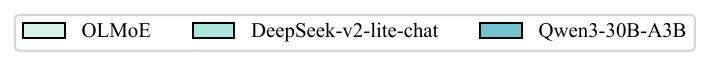}

    \begin{subfigure}[t]{0.48\textwidth}
      \centering
      \includegraphics[width=\linewidth]{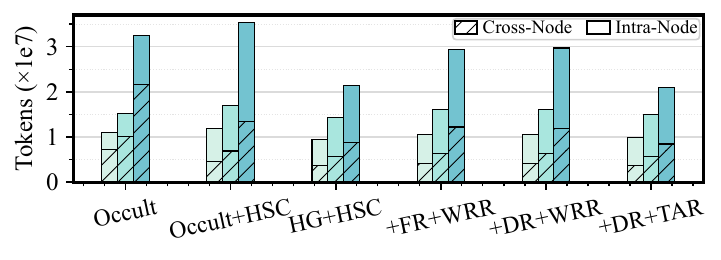}
      \caption{Communication traffic}
      \label{fig:rq_comm_tokens}
    \end{subfigure}
    \hfill
    \begin{subfigure}[t]{0.48\textwidth}
      \centering
      \includegraphics[width=\linewidth]{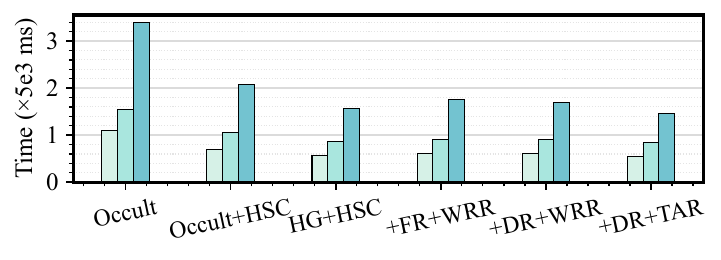}
      \caption{All-to-All time}
      \label{fig:rq_a2a}
    \end{subfigure}

    \begin{subfigure}[t]{0.48\textwidth}
      \centering
      \includegraphics[width=\linewidth]{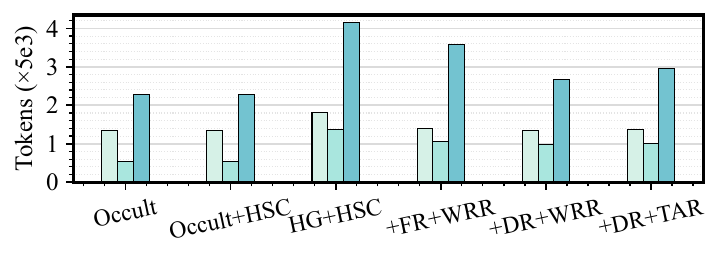}
      \caption{Mean per-layer standard deviation of GPU load}
      \label{fig:rq_loadstd}
    \end{subfigure}
    \hfill
    \begin{subfigure}[t]{0.48\textwidth}
      \centering
      \includegraphics[width=\linewidth]{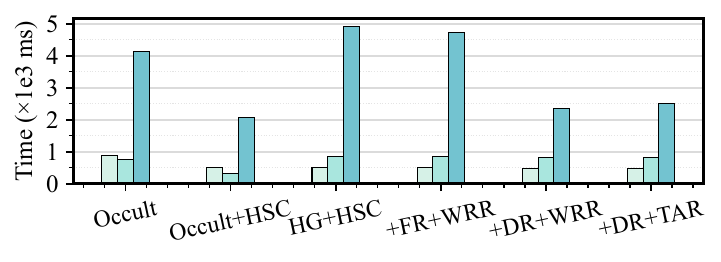}
      \caption{GPU idle time}
      \label{fig:rq_idle}
    \end{subfigure}
    \caption{\textbf{Component breakdown of system metrics.} Absolute values of communication overhead and computational load balance metrics corresponding to Table~\ref{tab:component_analysis}, evaluated across three models on the WikiText-2-v1 dataset.}
    \label{fig:component_abs}
\end{figure*}

To complement the analysis in Sec.~\ref{sec:component}, we visualize the absolute values of key communication overhead and computational load balance metrics under different component configurations.
Figure~\ref{fig:component_abs} provides a more intuitive view of how hierarchical sparse communication, non-uniform hierarchical grouping, dynamic replication, and locality-aware routing affect these metrics.
The observed trends are consistent with the relative comparisons reported in Table~\ref{tab:component_analysis}.

\end{document}